\documentclass[journal=apchd5,manuscript=letter]{achemso}
\usepackage{epstopdf}
\usepackage{amsmath}
\usepackage{bm}
\usepackage{braket}

\author{Josep Planelles}
\affiliation{Departament de Qu\'{\i}mica F\'{\i}sica i Anal\'{\i}tica,
Universitat Jaume I, E-12080, Castell\'o de la Plana, Spain}

\author{Alexander W. Achtstein }
\affiliation[Technical University of Berlin]{Institute of Optics and Atomic Physics, Technical University of Berlin, Strasse des 17. Juni 135, 10623 Berlin, Germany}
\email{achtstein@tu-berlin.de}

\author{Riccardo Scott}
\affiliation[Technical University of Berlin]{Institute of Optics and Atomic Physics, Technical University of Berlin, Strasse des 17. Juni 135, 10623 Berlin, Germany}

\author{Nina Owschimikow}
\affiliation[Technical University of Berlin]{Institute of Optics and Atomic Physics, Technical University of Berlin, Strasse des 17. Juni 135, 10623 Berlin, Germany}

\author{Juan I. Climente}
\affiliation{Departament de Qu\'{\i}mica F\'{\i}sica i Anal\'{\i}tica,
Universitat Jaume I, E-12080, Castell\'o de la Plana, Spain}
\email{climente@uji.es}

\title{Tuning intraband and interband transition rates via
	excitonic correlation in low-dimensional semiconductors}


\begin{document}

\begin{tocentry}
\includegraphics[height=3.5cm]{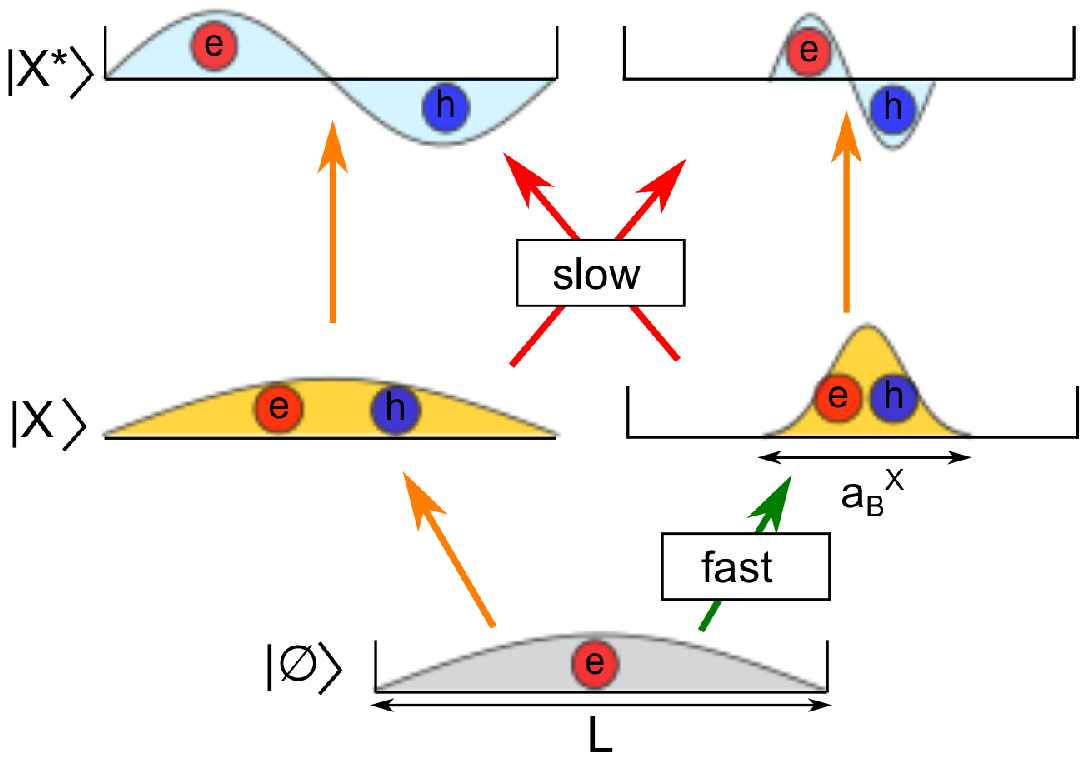}
\end{tocentry}

\begin{abstract}
We show that electron-hole correlation can be used to tune interband and intraband optical transition rates in semiconductor nanostructures with at least one weakly confined direction. The valence-to-conduction band transition rate can be enhanced by a factor $(L/a_B)^{N}$ -- with $L$ the length of the weakly confined direction, $a_B$ the exciton Bohr radius and $N$ the dimensionality of the nanostructure -- while the rate of intraband and inter-valence-band transitions can be slowed down by the inverse factor, $(a_B/L)^{N}$. Adding a hitherto underexplored degree of freedom to engineer excitonic transition rates, this size dependence is of interest for various opto-electronic applications. It also offers an interpretation of the superlinear volume scaling of two-photon absorption (TPA) cross-section recently reported for CdSe nanoplatelets, thus laying foundations to obtain unprecedented TPA cross sections, well above those of conventional two-photon absorbers. Further, our concept explains the background of the validity of the universal continuum absorption approach for the determination of particle concentrations via the intrinsic absorption. Potential applications of our approach include low excitation intensity confocal two-photon imaging, two-photon autocorrelation and cross correlation with much higher sensitivity and unprecedented temporal resolution as well as TPA based optical stabilization and optimizing of inter-subband transition rates in quantum cascade lasers.
\end{abstract}



Radiative transitions between the valence band (VB) and the conduction band (CB) in semiconductor quantum wells, wires and dots have been thoroughly investigated
over the last decades because of their potential for optical and opto-electronic applications.\cite{Nolte1999,Panfil2018,Tartakovskii2012}
 In these structures, the interband radiative rate is sensitive to electron-hole pair (exciton) interaction and quantum confinement.\cite{Kayanuma1988} Recent experiments with colloidal nanoplatelets (NPLs)\cite{Ithurria2011,Achtstein2012} and weakly confined quantum dots\cite{Tighineanu2016} have revealed band edge transition rates reaching 10 ns$^{-1}$ -- about two orders of magnitude faster than in strongly confined dots. This fast transition is a consequence of the so-called giant oscillator strength (GOST) effect, predicted for quantum wells\cite{Feldmann1987} and microcrystallites\cite{Kayanuma1988}.
The GOST effect arises when a strongly interacting exciton is allowed to propagate coherently over a large distance, resulting in a recombination rate which scales proportionally to such a distance. The resulting fast and tunable exciton radiative recombination is, for instance, of interest for high quantum yield emitters or low threshold lasing.\cite{Gong2015,Grim2014}

Excitonic transitions within CB or VB have received comparatively little attention since non-radiative (phonon- or Auger-mediated) recombination or relaxation generally takes place on a much shorter time scale,\cite{Panfil2018,Yazdani2018,Klimov1999,Spoor2017} posing a serious obstacle to observe emission. However, excitonic intraband and inter-valence-band absorption is relevant for several optical phenomena, including induced absorption in transient absorption spectroscopy,\cite{Klimov1999,Shi2013,El-Ballouli2014} non-linear optical processes including TPA, where concatenated inter- and intraband transitions often take place,\cite{Fedorov1996,Padilha2007,Heckmann2017} the development of mid-infrared detectors\cite{Wehrenberg2002} and emitters such as inter-subband quantum cascade lasers, as well as photoluminescence and electroluminescence of doped semiconductor nanocrystals\cite{Park2018,Kroupa2017}.

The current understanding of exciton intraband transitions largely relies on theoretical models investigating the effect of nanostructure size and shape,\cite{Fedorov1996,Turkov2011,Feng2009,Allione2015,Achtstein2015} neglecting the effect of electron-hole (e-h) interaction. In general, this is a reasonable assumption for strongly confined quantum dots, but it misses significant energetic contributions in  nanostructures of higher dimensionality.\cite{Parks2013} Even in quantum dots it is important for bound-to-unbound excitations.\cite{Kuhn2014}

In this work we study the effect of exciton interaction on inter- and intra-band transition rates of semiconductor nanostructures with different dimensionality. We show that exciton correlation is responsible for the GOST effect in e-h (interband) excitations. However, on electron-electron and hole-hole (intraband) excitations the exciton correlation has the opposite effect -- intraband and inter-valence-band transition rates between strongly bound and weakly bound exciton states scale inversely proportional to the size of the nanostructure. We provide analytical expressions based on effective mass theory for cuboidal nanocrystals, which reveal the dependence of transition rates on the structure dimensions and exciton Bohr radius.

The comprehensive approach to exciton transition rates we provide enables the rational design of radiative processes. Size-tunability can be activated or deactivated upon demand by the choice of proper initial and final states. Our results are immediately transferable to all semiconductor materials where a single-band description of electrons and holes holds. Potential applications include quantum cascade lasers with tunable inter subband transition rates or carrier multiplication, where fast radiative intraband cooling of highly excited states is not desirable. In particular, our results offer an interpretation for recent experiments of two-photon absorption (TPA) in CdSe NPLs for transitions into the continuum, where an unexpected quadratic volume dependence of the TPA cross sections has been found.\cite{Scott2015} This unusual result, which can be utilized for the design of two-photon absorbers with outstanding performance, can be explained if the TPA takes place through intermediate and final states with strong exciton correlation. Further we are able to substantiate a nanocrystal concentration determination method by intrinsic absorption in the nanocrystal absorption continuum.


\section{Results and discussion}

\begin{figure}
	\centering
		\includegraphics{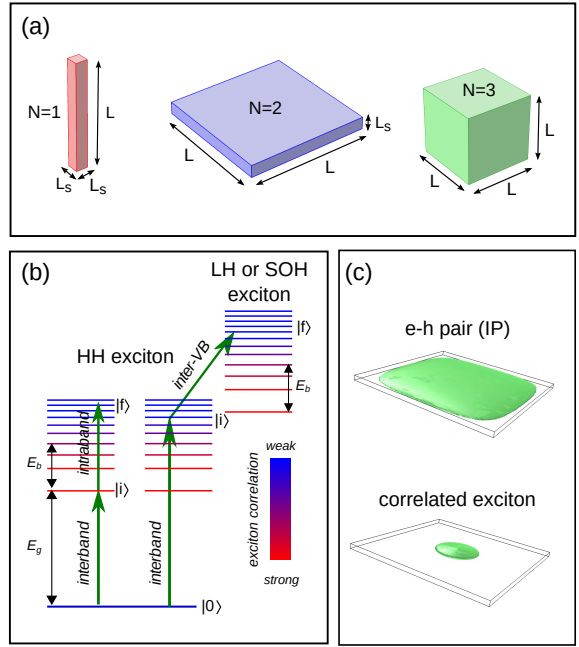}
	\caption{
(a) Cuboidal nanostructures under study. $N$ is the number of weakly confined dimensions of length $L$. 
(b) Different types of inter- and intraband transitions we consider. $|0\rangle$ is the state with all electrons in the valence band. 
 $E_{g}$ is the optical band gap energy and $E_b$ the exciton binding energy.
(c) Conditional probability of finding the electron after fixing the hole in the center of a quasi-2D NPL, corresponding to the ground state exciton without (top) and with (bottom) correlation factor. 
}\label{fig1}
\end{figure}

Our goal is to derive and analyze transition matrix elements for inter and intraband radiative transitions including exciton interaction within a single-band effective mass model. 
We consider cuboidal nanostructures as shown in Figure\,\ref{fig1}\,(a), with the length of the weakly confined directions  $L$ and that of the strongly confined ones $L_s$. 
The number of weakly confined directions ($N$) allows us to establish connections between our structures and quasi $N$-dimensional nanocrystals. Thus, cuboids with $N=1$ relate to nanorods, those with $N=2$ to nanoplatelets, and those with $N=3$ to bulky nanostructures e.g. large dots.

Figure\,\ref{fig1}\,(b)  schematically shows the different types of optical transitions we consider. Initially all electrons are in the valence band.  Within our formalism this is taken as the vacuum state $|0\rangle$ (no e-h pair, always uncorreleated). Next, a photon excites one electron across the band gap to form an e-h pair, $|i\rangle$, through an interband transition. Subsequent excitations to a final state $|f\rangle$ can be of intraband or interband character. For intraband transitions, a second photon (e.g. from a two-photon absorption process) excites the electron within the CB or the hole within its valence subband. Hole excitations between different valence subbands are also possible, e.g. heavy-hole (HH) to light-hole (LH) or split-off hole (SOH) . These are referred to as inter-valence-band (inter-VB) transitions.

The states $|i\rangle$ and $|f\rangle$ can vary from strongly bound excitons  (those near the optical band edge $E_g$) to unbound excitons (those well above the exciton binding energy, $E_g+E_b$ in figure \ref{fig1}(b)). We  consider both kind of states. 
 To this end, the envelope function of a general exciton state $|m\rangle$ is written as:\cite{Rajadell2017,Planelles2017} 
\begin{equation}
|f_m \rangle = N^m \, \Phi_{IP}^m(\mathbf{r}_e,\mathbf{r}_h) \, \Phi^m_{corr}(|\mathbf{r}_e-\mathbf{r}_h|).
\label{eq:Psicorr}
\end{equation}
\noindent Here, $N^m$ is the normalization constant and $\Phi_{IP}^m$ the non-normalized independent particle (IP) envelope function, which we take as a Hartree product of electron and hole particle-in-box functions: $\Phi_{IP}^m = \prod_\alpha f_{n_\alpha^e}(\alpha_e) f_{n_\alpha^h}(\alpha_h)$, with $\alpha=x,y,z$ and $f_{n_\alpha} = \cos{(n_\alpha\, \pi\, \alpha/L_\alpha)}$ for odd quantum number $n_\alpha$, or $f_{n_\alpha} = \sin{(n_\alpha\, \pi\, \alpha/L_\alpha)}$ for even $n_\alpha$. $\Phi^m_{corr} = e^{-r_{eh}/a_B^m}$ is a Slater correlation factor, with $r_{eh}$ the e-h separation and $a_B^m$ the effective Bohr radius of state $m$ along the weakly confined directions.  For weakly bound states ($a_B^m \rightarrow L$, $\Phi^m_{corr} \rightarrow 1$) Eq.~(\ref{eq:Psicorr}) simplifies to an independent e-h wave function. However, strongly interacting excitons have radii $a_B^m \ll L$ and then the wavefunction $|m\rangle$ departs significantly from that of independent particles (IPs).

To obtain explicit matrix elements, in the following, we choose the case of a cuboid with $N=2$ weakly confined dimensions (nanoplatelet). Other dimensionalities behave in a similar way, as can be seen in the Supporting Information (SI). The correlation factor can then be simplified to an in-plane expression $\Phi^m_{corr} \approx e^{-r_{eh,\perp}/a_B^m}$, with $r_{eh,\perp}=\sqrt{(\mathbf{r}_{e,\perp} - \mathbf{r}_{h,\perp})^2}$. 
For a squared NPL in the strong confinement limit ($a_B^m \rightarrow L$), the normalization factor is just:
\begin{equation}
N^m = \frac{8}{L L_s}\,\frac{1}{L} = N_{IP}^m.
\label{eq:NIP}
\end{equation}
\noindent In the strong correlation limit ($a_B^m \ll L$), the normalization factor can be approximated as:
\begin{equation}
N^m= \frac{8}{L L_s} \,\frac{1}{a_B^m}\,\frac{\beta}{\sqrt{\pi}} = N_{corr}^m,
\label{eq:Ncorr}
\end{equation}
\noindent where $\beta$ is a constant that depends on the IP quantum numbers (for the ground state $\beta=\sqrt{8}/{3}$).\cite{Planelles2017} The replacement of a $1/L$ term in $N_{IP}^m$ by $1/a_B^m$ in $N_{corr}^m$ reflects the fact that  correlated excitons concentrate in a much smaller region of space than the NPL boundaries, as illustrated in Figure\,\ref{fig1}\,(c).

\subsection{Interband transitions}

With the expressions above we can analyse the interband transition matrix element. Transitions between VB and CB states (e-h transitions) are from the vacuum state $|0\rangle$ to a state $|i\rangle$. The dipole matrix element is proportional to the e-h envelope function overlap:
\begin{equation}
S_{eh} = \langle f_0 | \delta(\mathbf{r}_e-\mathbf{r}_h) | f_i \rangle = 
N^i \, \left( \frac{L}{2} \right)^2\,\left( \frac{L_s}{2} \right)
\delta_{n_x^e,n_x^h}\,
\delta_{n_y^e,n_y^h}\,
\delta_{n_z^e,n_z^h}.
\label{eq:Seh}
\end{equation}
\noindent It can bee seen that the only influence of exciton correlation in Eq.~(\ref{eq:Seh}) is to change $N^i$, since the correlation factor vanishes when $\mathbf{r}_e=\mathbf{r}_h$. In the case of an independent e-h we use $N^i=N_{IP}^i$ and find that the interband matrix element for allowed transitions is size independent $S_{eh}=1$. For a correlated exciton ($N^i=N_{corr}^i$) it becomes $S_{eh}=\beta\,\sqrt{A/A_X^i}$, introducing a size sensitivity, where $A=L^2$ is the NPL area and $A_X^i=\pi (a_B^i)^2$ the exciton area. The size insensitivity of interband matrix element in the IP case has been confirmed in strongly confined quantum dots\cite{vanDriel2005}. There the correlation vanishes due to the strong isotropic confinement. In turn, the rapid increase with the NPL area in the correlated case is a manifestation of the GOST effect for strongly interacting excitons in the $L \gg a_B$ regime.\cite{Kayanuma1988,Ithurria2011,Achtstein2012,Tighineanu2016,Feldmann1987}

Another case of interest are inter-VB transitions, where an exciton in state $|i\rangle$ changes to an exciton in state $|f\rangle$ e.g. by promoting a hole from the HH to the LH or SO subbands. These are hole-hole transitions, but they are also affected by interactions with electrons. If the transition takes place between two states of noninteracting e-h pairs, the envelope part of the matrix element is simply: 
\begin{equation}
\langle f_i | f_f \rangle = 
\prod_{\alpha=x,y,z} \delta_{(n_{\alpha}^e)_i,(n_{\alpha}^e)_f}\, \delta_{(n_{\alpha}^h)_i,(n_{\alpha}^h)_f}
\label{eq:inter}
\end{equation}
\noindent which shows that the transition is only allowed for identical IP quantum numbers 
$(n_\alpha^j)_i=(n_\alpha^j)^f$, and independent of the nanocrystal size. 
To illustrate the effect of e-h interaction, we also restrict to the case of identical IP quantum numbers, because these are expected to yield the largest matrix elements. For an inter-VB transition between two correlated exciton states, in the limit of $a_B \ll L$, we obtain (see SI):
\begin{equation}
\langle f_i | f_f \rangle =  4\, \frac{a_B^i \, a_B^f}{\left(a_B^i + a_B^f\right)^2}
\label{eq:intercorrcorr}
\end{equation}
while for an inter-VB transition between a correlated exciton state $|i\rangle$ and an uncorrelated e-h pair state $|f\rangle$, it becomes: 
\begin{equation}
\langle f_i | f_f \rangle = 
\sqrt{18\pi} \, \frac{a_B^i}{L}.
\label{eq:interIPcorr}
\end{equation}
Eqs.~(\ref{eq:intercorrcorr}) and (\ref{eq:interIPcorr}) show no trace of the GOST effect observed for valence-to-conduction band transitions. Equation~(\ref{eq:intercorrcorr}) shows that inter-VB transitions between two correlated exciton states have no explicit size dependence. In fact, for identical correlation ($a_B^i=a_B^f$) one obtains $\langle i | f \rangle=1$, so that there is no size dependence at all, as in the IP case. In turn, Eq.~(\ref{eq:interIPcorr}) reveals that transitions between correlated exciton and noninteracting e-h states not only show no GOST effect, but rather the inverse effect, since an extra $a_B^i/L$ factor appears in the matrix element. In other words, while exciton interaction makes the e-h interband transition rate (squared matrix element) increase with the NPL area as $A/A_X$,  it also makes inter-VB transition rates decrease as $A_X/A$.

\subsection{Intraband transitions}

The general form of the intraband matrix element in our formalism is:
\begin{equation}
\langle i | \mathbf{p} | f \rangle = 
\langle f_i | \mathbf{p} | f_f \rangle = 
N^i\, N^f \, \langle \Phi_{IP}^i \, \Phi_{corr}^i | \mathbf{p} | \Phi_{IP}^f\, \Phi_{corr}^f \rangle,
\end{equation}

\noindent where the linear momentum operator $\mathbf{p}=\mathbf{p}^e+\mathbf{p}^h$ can act either on electron or hole coordinates. We consider the $p_{x}^e$ component (the others behave similarly) and start with the case of transitions between two IP states, as is the case in strongly confined nanostructures.  In this case $\langle i | p_x^e | f \rangle = N_{IP}^i\, N_{IP}^f \, \langle \Phi_{IP}^i | p_x^e | \Phi_{IP}^f \rangle$. Functions' orthogonality restricts the accessible excited states to excitations of the $x_e$ function and, in particular, to those having different parity:
%
\begin{equation}
\langle i | p_x^e | f \rangle = -i \, \hbar\; g\left[(n_{x}^e)_i,(n_{x}^e)_f\right] \; \delta_{(n_{x}^h)_i,(n_{x}^h)_f} \,
\prod_{\alpha=y,z} \delta_{(n_{\alpha}^e)_i,(n_{\alpha}^e)_f}\, \delta_{(n_{\alpha}^h)_i,(n_{\alpha}^h)_f}.
\label{eq:intra}
\end{equation}
\noindent with:
\begin{equation}
g\left[n_i,n_f\right] = 
\begin{cases}
\frac{4}{\pi}\,n_i \,\frac{(-1)^{(n_f-n_i-1)/2}}{n_f^2-n_i^2}\, k_f & \mbox{if }  n_i+n_f \mbox{ is odd},\\
0 & \mbox{if } n_i+n_f \mbox{ is even},
\end{cases}
\label{eq:g}
\end{equation}
\noindent where $k_f=n_f\pi/L$. 
From the denominator in Eq.~(\ref{eq:g}), it is clear that $g\left[n_i,n_f\right]$, and hence intraband matrix elements, peak at transitions between consecutive states, $n_i = n_f \pm 1$. We shall see later that this quasi-selection rule has implications for non-linear optical processes. If we restrict to such transitions, it is easy to see that
 the matrix element increases with $n_f$, which indicates that transition rates between high-energy levels are faster than those between low-energy ones.

The size dependence of the intraband matrix element depends on the nature of the experiment. If the transition takes place between two fixed levels, $|i\rangle$ and $|f\rangle$,  the $1/L$ factor in $k_f$ implies that the transition becomes less likely with increasing size. A more common scenario in experiments, however, is that $|i\rangle$ is the ground state and $|f\rangle$ an excited state -- set by the resonance condition with a photon of fixed energy -- whose precise quantum number varies with the nanostructure size. 
In this case it is easy to check numerically that, for large photon energies $h \nu$,  $\hbar^2 k_f^2 /2\mu\pi^2 \approx h\nu - E_g$, with $\mu$ the exciton reduced mass.
It follows that $k_f \propto L^0$, 
and the matrix element is roughly size-independent.  This is relevant for high energy transitions, which we will discuss in the next section.
If the photon energy is small, instead, the matrix element becomes largest for the $L$ values such that the resonant level fulfills $ n_f \approx n_i \pm 1$.


We next consider the case of transitions between two correlated exciton states. Here, $\langle i | p_x^e | f \rangle = N_{corr}^i\, N_{corr}^f \, \langle \Phi_{IP}^i \Phi_{corr}^i | p_x^e | \Phi_{IP}^f \Phi_{corr}^f \rangle$. Assuming $\Phi_{corr}^i$ has even parity in all directions (even quantum numbers $n_\alpha^e$ and $n_\alpha^h$), and the final state odd parity in the $x_e$ direction, the dipole-allowed element is proportional to (see SI):
\begin{equation}
\langle i | p_x^e | f \rangle \propto g\left[(n_{x}^e)_i,(n_{x}^e)_f\right]\, \frac{a_B^i a_B^f}{\left(a_B^i+a_B^f\right)^2}.
\label{eq:intracorrcorr}
\end{equation}
\noindent For a transition from correlated exciton to IP e-h states the matrix element becomes: 
\begin{equation}
\langle i | p_x^e | f \rangle \propto g\left[(n_{x}^e)_i,(n_{x}^e)_f\right]\, \frac{a_B^i}{L}.
\label{eq:intraIPcorr}
\end{equation}
\noindent Eqs.~(\ref{eq:intra}), (\ref{eq:intracorrcorr}) and (\ref{eq:intraIPcorr}) are reminiscent of their inter-VB counterparts,  Eqs.~(\ref{eq:inter}), (\ref{eq:intercorrcorr}) and (\ref{eq:interIPcorr}), respectively. Thus, if initial and final states have a similar degree of exciton correlation, the matrix element is size independent. If the difference is large, however, an inverse GOST effect takes place, which suppresses the transition rate proportional to $A_x/A$. This is especially relevant for the design of quantum cascade lasers with efficient intraband transitions that lead to higher gain. Instead of taking hetero-structures and superlattices of virtually infinite lateral size, column like structures with finite cross-sectional area may allow superior performance.

A qualitative interpretation of our findings on the dimensional scaling so far is as follows: Radiative transitions between the VB and CB band imply e-h recombination.  Compared to the IP case, excitonic interaction keeps the two particles close together and enhances the transition rate, leading to the GOST effect.  Intraband and inter-VB processes, by contrast, imply electron-electron or hole-hole transitions. In transitions from correlated to IP states, the strong exciton interaction of the initial state confines the electron in a small region near the hole with area $A_X$,  reducing its overlap with the electron wave function of the final state, which is delocalized over the entire NPL area $A$.  This eventually results in the $\sqrt{A_X/A}$ factor of Eqs.~(\ref{eq:interIPcorr}) and (\ref{eq:intraIPcorr}). If both states are correlated or both uncorrelated, however, the overlap tends to $\sqrt{A_X/A_X}$ or $\sqrt{A/A}$, respectively,  so that there is no additional scaling with size.

Table\,\ref{tab1} summarizes our results for nanostructures with $N$ weakly confined directions for inter, inter-VB and intra band transitions. In general, exciton correlation introduces a different size dependence on inter- and intraband transition matrix elements, transitions between strongly and weakly correlated exciton states become particularly sensitive to the size $L$ of the nanostructure in the weakly confined directions. They benefit from the GOST (interband) and inverse GOST (intraband, inter-VB) effects,  which can be used to tailor the transition rates. 
\begin{table}
\begin{tabular}{ c c c c }
\hline
correlation   	                &  $\langle i | p_\alpha| 0\rangle_{inter}$ 	&  $\langle f | p_\alpha | i\rangle_{\mbox{{\scriptsize inter-VB}}}$  & $\langle f | p_\alpha | i\rangle_{\mbox{{\scriptsize intra}}}$ \\
\hline
weak         $\leftrightarrow$  weak	&  $L^0$ 	& 		$L^0$				& $q\,k_f $	\\[5pt]
strong       $\leftrightarrow$  weak	&  $\left( \frac{L}{a_B^i} \right)^{N/2}$ &  $\left( \frac{a_B^i}{L} \right)^{N/2}$			&  $q\,k_f\,\left(\frac{a_B^i}{L}\right)^{N/2}$ \\[5pt]
strong       $\leftrightarrow$  strong &  $-$		& 			$\mu^N$			& $q\,k_f \,\mu^N$  \\[5pt] 
\hline
\end{tabular}
\caption{Size and quantum number dependence of dipole allowed transition matrix elements in cuboidal nanostructures with $N$ weakly confined directions of length $L$.  The matrix element depends on the strength of exciton correlation in the initial and final states, as indicated in the first column. Weakly and not correlated (IP) states, including the vacuum state, result in identical entries.
$q=n_i/(n_f^2-n_i^2)$, $\mu = (a_B^i\,a_B^f)^{1/2}/(a_B^i+a_B^f)$, $k_f = n_f \pi /L$. 
}\label{tab1}
\end{table}

\subsection{Implications on linear absorption and TPA}

In this section we study how our findings above affect the linear absorption at high energies and the TPA processes based on examples for nanoparticle systems.

\begin{figure}
\centering
\includegraphics{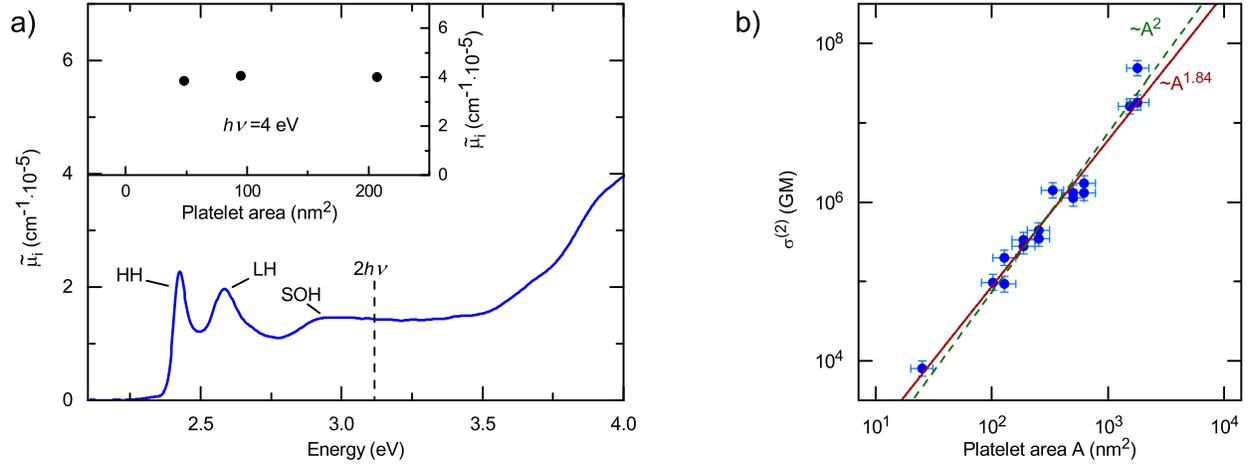}
\caption{ (a) Linear absorption of an exemplary 4.5\,ML CdSe NPL with lateral sizes of 19x5\,nm$^2$. The black dashed line  indicates the two photon energy $2h\nu$ of 3.1\,eV.  Heavy hole (HH), light hole (LH) and split off (SOH) exciton transitions are indicated. Each band has strongly correlated lowest exciton states (indicated maxima), a quasi continuum  of weakly correlated states and free electron-hole pair states in the continuum (referred as IP states in the text) as indicated in figure\,\ref{fig1}\,(b). Inset: Intrinsic absorption at 4\,eV for 4.5\,ML platelets of different area. The second datapoint belongs to the platelet in (a). Data from Ref.\citenum{Achtstein2015}. (b) Logarithmic plot of area dependence of the TPA cross section measured at 800\,nm (1.55\,eV) for CdSe NPLs of 3.5, 4.5 and 5.5\,ML thickness and varying lateral dimensions. Points are experimental values from Ref.\,\citenum{Scott2015}. Solid line: power law fit. Dashed line: Forced square dependence fit.}
\label{TPAvol}
\end{figure}

 In Ref.\,\citenum{Achtstein2015a} it has been shown experimentally that the linear absorption of CdSe nanoplatelets, or more precisely the intrinsic absorption $\tilde{\mu}_i=\sigma/V$, high in  the continuum (e.g. at 4\,eV) is an universal quantity for a semiconductor material. The inset of Figure\,\ref{TPAvol}\,(a) shows that, indeed,  the intrinsic absorption at 4\,eV remains nearly constant for a series of different laterals sizes (areas) of the nanoplatelets at fixed thickness of 4.5\,ML.  Similar findings have been obtained for semiconductor quantum dots and nanorods\cite{Achtstein2013,Capek2010,Kamal2012}, and they enable accurate concentration determination using effective media and local field theory.
However, to our knowledge, the reason for this universal volume scaling has remained unclear.

Our model for the dimensional scaling of inter band transitions provides a direct explanation. 
For example, in the case of NPLs, at 4\,eV the linear absorption process occurs into a continuum state\cite{Grim2014}, i.e. we observe a transition from a uncorrelated crystal ground-state $|0\rangle$ to an uncorrelated continuum state $|i\rangle$. This is evidenced by the findings of Scott et al.\cite{Scott2017}, who showed that the absorption above $3.1$ eV is spatially isotropic, and hence HH, LH and SO continua all contribute to the absorption, as only in this case the HH, LH and SO valence band Bloch functions add up to an isotropic dipole distribution.  
According to Table \ref{tab1}, the matrix element $\langle i | p_\alpha | 0\rangle$ for a continuum transition has an $L^0$ dependence so that the linear absorption cross section, from first order Fermi's Golden Rule, is $\sigma_{1PA}\propto f_{LF}^2\, W_{1PA} \propto (2\pi f_{LF}^2/\hbar) \left|\langle i | \mathbf{p} | 0\rangle  \right| ^2 \rho( h\nu_{i} - h \nu )$, with $\rho( h\nu_{i} - h \nu )\propto L^N$ the density of final states, which for a NPL scales with the area $A=L^2$ (or volume for fixed quantum well thickness) in a (quasi) 2D system. $f_{LF}$ is a geometry dependent local field factor. Therefore the absorption cross section per unit volume $\tilde{\mu_i}=\sigma/V=\sigma/AL_s$ high in the absorption continuum is a universal quantity independent on $A$. It varies only weakly  with $f_{LF}$ for a given semiconductor nanomaterial with the particle shape,\cite{Achtstein2015a} and is constant if the nanostructure's shape is maintained and only its size is altered.
Comparable arguments hold for nanorod and quantum dot materials, where a similar continuum intrinsic absorption scaling has been observed,\cite{Achtstein2013,Capek2010,Kamal2012}
as also expected from our model.  This is a first demonstration of the validity and impact of our approach.


Next we concentrate on the implications of dimension scaling for two-photon absorption and the near quadratic TPA scaling observed for nanoplatelets.
For comparison with experiments, we focus again on CdSe NPLs, but the conclusions are general for quasi-$N$-dimensional cuboids. We showed in recent experiments that for two-photon energies of 3.1\,eV (two photons of 800\,nm) the TPA cross-section ($\sigma^{(2)}$) of these nanoplatelets scales almost quadratically with the NPL volume.\cite{Scott2015} In sharp contrast to the approximately linear volume scaling reported in quantum dots and dot-based structures,\cite{Pu2006,Allione2015,Achtstein2013,Blanton1994,Feng2009} this translates into extraordinarily high $\sigma^{(2)}$ values (up to over 10$^7$ GM), making platelet-like nanostructures optimal candidates for two-photon imaging,  nonlinear optoelectronics and even two-photon autocorrelation. This raises the question of the origin of this unusual power-law behavior, whether it is a unique property of CdSe NPLs and whether it can be generalized to other systems or dimensionalities.

Figure~\ref{TPAvol}\,(b) shows that the cross section of CdSe NPLs scales quadratically not only with the particle volume\cite{Scott2015} ($V=AL_s$), but also with its area $A$, as the changes in NPL volume are mainly given by changes in the area (approx. two orders in magnitude), while the changes in thickness from 3.5 to 5.5\,ML are minor. A power function fit yields an $A^{1.84\pm 0.09}$  dependency (solid line), where most data points meet a quadratic fitcurve (dashed line) within their error bars.

To understand this behavior, we consider the TPA cross section of a particle. It can be evaluated as\cite{BoydNLO}
$\sigma^{(2)} = {W_{TPA}}/{I_{ph}^2}$,
%
where $I_{ph}$ the photon flux density 
and $W_{TPA}$ is the two-photon transition rate at a laser energy $h \nu$:
\begin{equation}
W_{TPA} (h\nu) = \frac{2\pi}{\hbar} \sum_{f}  \left| \sum_{i} \frac{  \langle f | H' | i\rangle \langle i |H'| 0\rangle} { h\nu_i - h \nu} \right| ^2 \Gamma \left( h\nu_{f} - 2h \nu \right).
\label{eqW0}
\end{equation}
\noindent 
\noindent Here we have assumed the initial state $|0\rangle$ as the reference energy, $|i\rangle$ an intermediate state and $|f\rangle$ the final one. $\Gamma \left( h\nu_{f} - 2h \nu \right)$ is a Gaussian function which accounts for the energy resonance condition, considering our laser source is not monochromatic  but has a finite bandwidth. In dielectrically heterogeneous media -- as colloidal NPLs embedded in organic ligands and solvents -- the external field ($E_{ext}$) differs from the local field ($E$) inside the nanoparticle
and a local field factor $f_{LF}=E/E_{ext}$ arises such that 
$\sigma^{(2)} = f_{LF}^4\,{W_{TPA}}/{(I_{ph}^{L})^2}$,
\noindent where $I_{ph}^L$ is the local photon flux density. 
Ref.~\citenum{Scott2015} studied the influence of $f_{LF}$ on $\sigma^{(2)}$ and concluded it was not enough to explain the drastic variation of TPA crossection with NPL size in the experimental data. Thus, the characteristic behavior of NPLs must arise from the TPA rate, $W_{TPA}$. We shall focus our analysis on this sole factor.

\begin{figure}
        \centering
                \includegraphics{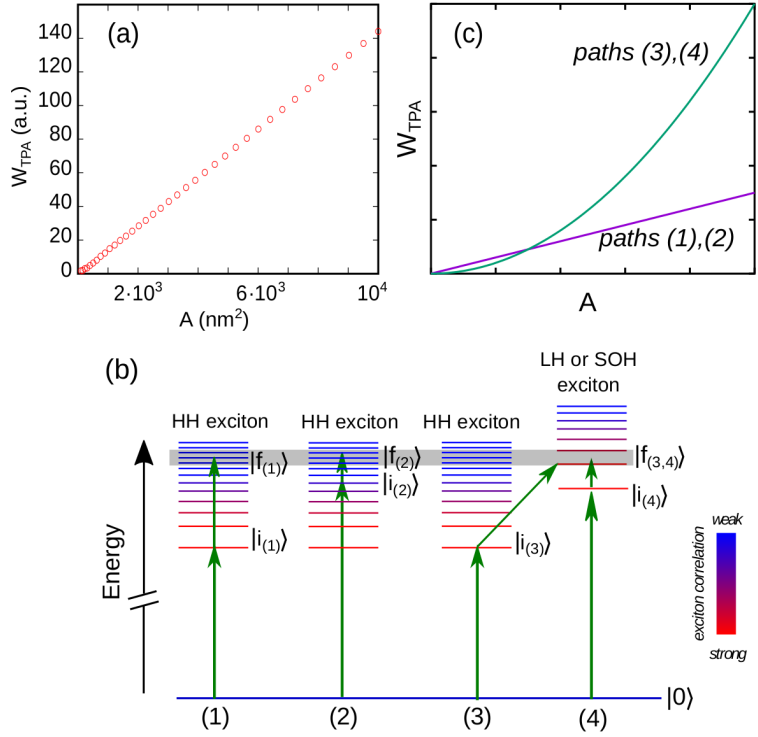}
        \caption{
		(a) TPA rate calculated within an IP model using Eq. \ref{eqW0}. CdSe electron and hole masses are taken from Ref.~\citenum{Norris1996}, $E_g=2.4$ eV and the laser bandwidth is set to $50$ meV. A clear linear dependence with the area is observed, in contrast with the experimental data. 
		(b) Diagram of possibly relevant paths in the TPA of CdSe NPLs under $800$ nm laser: 
	the final state lies near the two-photon energy (shaded region), and the correlation strength of intermediate 
	($|i\rangle_{(1-4)}$) and final states ($|f\rangle_{(1-4)}$) varies.
		(c) Schematic representation of the NPL area dependence of TPA cross-section for paths (1)-(4). 
 Quadratic area dependence arises only if both intermediate and final states are correlated excitons, as in paths (3) and (4).}
        \label{fig3}
\end{figure}

We start by calculating $W_{TPA}$ in the independent particle (IP) limit. 
Because experiments use sub-band gap photon energies ($h \nu =1.55$\,eV vs. $E_g \geq 2.25$\,eV  for our CdSe NPLs),  there is no quasi-resonant stationary intermediate state. 
It must rather be a virtual non-stationary state.  We then consider the contribution of a large number of eigenstates $|i \rangle$ to such a virtual level in Eq.~(\ref{eqW0}).
 For illustration, we assume the TPA process involves an interband transition --forming an electron and a HH-- followed by an intraband transition, 
and compute $W_{TPA}$ numerically with particle-in-the-box envelope functions and energies.
The resulting $W_{TPA}$ displays a clear linear dependence with the area, as shown in Fig.~\ref{fig3}\,(a). 
 Linear scaling is also obtained considering other IP excitations (e.g. two interband transitions, not shown). 
We then conclude an IP model cannot explain the quadratic cross-section scaling observed in the experiments.

At high two-photon energies, however, the resonant state is not necessarily an unbound (IP) e-h pair. 
It can also be a bound exciton state of an excited (LH, SOH) valence subband. 
This point is exemplified in Fig.~\ref{TPAvol}\,(a), which shows the linear absorption spectra of $4.5$\,ML NPLs. 
It can be seen that excitations around $3.1$ eV (as corresponding to our laser of 800 nm) 
fall in the regions of strongly correlated SOH exciton transitions\cite{Grim2014} as the exciton binding energies separating the lowest exciton state from the respective continuum in each band are of the order of $E_b\sim 170-260$\,meV.\cite{Naeem2015,Scott2016a} 
Additionally there is an underlying background from the HH and LH continua. 
We remark that the laser linewidth of $\sim50\,$meV in Eq. \ref{eqW0} is high enough to assume the final state to be a continuum of states, as for example in an ideal 2D system the exciton binding energies of the still strongly correlated second and third excited states are $1/9 \, E_b$ and $1/25\,E_b$.\cite{Basu97} Hence as the binding energy of the second and third state is $E\sim 24$\,meV and $E\sim 8$\,meV, the laser can address the whole quasicontinuum of states with half of its FWHM so that there is density of final states, as indicated in Equation \ref{eqW0} (and later in \ref{eq:TPA2}).

Calculations involving correlated exciton ground and excited states in NPLs are extremely demanding,  as there are no simple envelope wave functions except for the low-lying states.\cite{Rajadell2017,Richter2017}  To circumvent this peculiarity, we resort to a more intuitive, qualitative analysis of $W_{TPA}$ based on the size dependences of Table\,\ref{tab1}. 
 We consider the four representative types of TPA processes shown in Figure\, \ref{fig3}\,(b). 
Paths (1) and (2) reach a final unbound exciton state via an interband transition followed by an intraband transition, over a correlated (path (1)) or uncorrelated (path (2)) intermediate state. Paths (3) and (4) reach a bound exciton by an interband transition to a correlated intermediate state followed by an inter-VB (path(3)) or intraband transition (path(4)).

%

Table\,\ref{tab2} shows the size dependence for the matrix elements of the different TPA paths of Fig.~\ref{fig3}\,(b), as obtained from Table\,\ref{tab1} with $k_f \propto L^0$. The latter condition follows from the resonance condition in Eq.~(\ref{eqW0}), since the two-photon energy was fixed for all experimental measurements in Fig.~\ref{TPAvol}\,(a). If TPA takes place through paths (1) or (2), where the final state is an unbound HH exciton, the product of inter and intraband matrix elements leads to a null dependence of the matrix element product on the size, $L^0$. 
A net size dependence of the product of the squared matrix elements can only be obtained, if both intermediate and final states are bound (correlated) excitons, as in paths (3) and (4).
\\
\begin{table}
\begin{tabular}{ c c c c c }
\hline
path	&  $\langle i | H' | 0 \rangle$  &  $\langle f | H' | i \rangle$ &  $\rho(h \nu_f)$   &   $W_{TPA}$ \\ 
\hline
1	& $\left( \frac{L}{a_B^i} \right)^{N/2}$ & $\left(\frac{a_B^i}{L}\right)^{N/2}$	& $L^N$	&  $L^N$ \\ [5pt]
2	& $L^0$ 				 & $L^0$ 				& $L^N$	&  $L^N$ \\ [5pt]
3	& $\left( \frac{L}{a_B^i} \right)^{N/2}$ & $L^0$				& $L^N$	&  $\left(L^2\,\frac{1}{a_B^i}\right)^{N}$ \\ [5pt]
4	& $\left( \frac{L}{a_B^i} \right)^{N/2}$ & $L^0\,\left(\frac{a_B^i\,a_B^f}{(a_B^i+a_B^f)^2}\right)^{N/2}$	& $L^N$	&  $\left(L^2\,\frac{a_B^i\,a_B^f}{a_B^i\,(a_B^i+a_B^f)^2}\right)^N$ \\ [5pt]
\hline
\end{tabular}
\caption{Size dependence of transition matrix elements, density of states (DOS) and TPA rate ($W_{TPA}$) in cuboidal nanostructures with $N$ weakly confined directions of length $L$, for different TPA paths, see Fig.~\ref{fig3}\,(b).}\label{tab2}
\end{table}
 
The next term we need to consider to understand $W_{TPA}$ is the density of states (DOS). In Eq.~(\ref{eqW0}) the sum over intermediate and final states involves 12 degrees of freedom ($(n_{\alpha}^j)_m$, with $j=e,h$, $\alpha=x,y,z$ and $m=i,f$). However, selection rules reduce this number.\cite{selrules}  The interband transition $\langle i | H' | 0 \rangle$ imposes $(n_{\alpha}^e)_i = (n_{\alpha}^h)_i$, see Eq.~(\ref{eq:Seh}). If the subsequent transition $\langle f | H' | i \rangle$ is also an interband process, the matrix element gives $(n_{\alpha}^j)_i = (n_{\alpha}^j)_f$, as seen in Eq.~(\ref{eq:inter}). If it is an intraband process instead,  Eq.~(\ref{eq:intra}) imposes $(n_{\alpha}^j)_i = (n_{\alpha}^j)_f$ as well, except for the direction and carrier acted upon by $\mathbf{p}$. Here we have a quasi-selection rule $n_i = n_f \pm 1$, as previously noted from Eq.~(\ref{eq:g}). All restrictions considered, $W_{TPA}$ involves only the sum over 3 degrees of freedom, namely the IP quantum numbers of electron or hole final states $(n_{\alpha}^j)_f$. For a given TPA path $\lambda$, we can then write:
\begin{equation}
W_{TPA}^\lambda (h\nu) 
\propto
\left| 
\frac{  \langle f_{\mbox{{\tiny ($\lambda$)}}} | H' | i_{\mbox{{\tiny ($\lambda$)}}}\rangle \langle i_{\mbox{{\tiny ($\lambda$)}}} |H'| 0\rangle} { h\nu_{i_{\mbox{{\tiny ($\lambda$)}}}} - h \nu} 
\right| ^2 \rho ( h\nu_{f} ).
\label{eq:TPA2}
\end{equation}
%
\noindent Here we have replaced the sum over final states by the joint DOS $\rho ( h \nu_f )$, which in cuboidal nanostructures is determined by the dimensionality $N$ as $\rho_{ND}\propto L^N$. We have also considered that, as discussed above, for a given final state $|f_{(\lambda)}\rangle$ within the laser bandwidth, there is only one (two, in the case of intraband transitions) relevant intermediate states, $|i_{(\lambda)}\rangle$.
From Eq.~(\ref{eq:TPA2}), we can extract the net size dependence of $W_{TPA}$ for each path.  The result is shown in the last column of Table\,\ref{tab2}.  By comparison with the experiments of NPLs ($N=2$), we conclude that only paths involving intermediate and final  states with strong exciton correlation, i.e. paths (3) and (4), explain the superlinear dependence observed in Fig.~\ref{TPAvol}\,(a), as $W_{TPA} \propto A^2/A_X$. If any of the two states lacks correlation, $W_{TPA} \propto A$, which agrees with our numerical calculations of Fig.~\ref{fig3}\,(a).\cite{agree} 
These results are summarized schematically in Fig.~\ref{fig3}\,(c) for the different paths (1 - 4).
 
One can argue that in actual NPLs,  $W_{TPA}$ will gather contributions from different paths, with both linear ($\lambda=1,2$) and quadratic ($\lambda=3,4$) size dependence.
Yet, because NPLs have large areas, paths (3) and (4) --scaling with $A^2$ instead of $A$-- are expected to prevail.
 Further, path (1) is unlikely due to the quasi selection rule $n_i=n_f\pm 1$ imposed by Eq.(\ref{eq:g}).
 Path (2) is also less likely, because the oscillator strength $f_o$ of the first (interband) transition is expected to be low, as in a 2D system the transition oscillator strength for a $|0 \rangle \rightarrow |i \rangle $ scales with $f_o\propto (n+1/2)^{-3}$ ($n=0,1,2 \ldots$)\cite{Basu97}, so that the transition oscillator strength to a higher excited state state in the first transition is low. 
On the other hand, if the final state is correlated paths -- as in paths (3,4)--, it is not by chance that so is the intermediate state.  If the transition is through an intraband process, it most likely implies two adjacent levels, see Eq.~(\ref{eq:g}). If it is through an inter-VB transition, final and intermediate states must have the same quantum numbers, see Eq.~(\ref{eq:inter}), so that exciton states are analogous but in different subbands. 

A further argument for the identification of the $A^2$ scaling interpretation via paths (3) and (4) comes from $k$-space resolved two-photon spectroscopy. 
It has been measured that 85\% of the transition dipoles for TPA lie in-plane oriented with respect to the CdSe nanoplatelets.\cite{Heckmann2017} 
Since the Bloch function symmetries of the involved HH, LH, SOH and electron bands and the envelope functions are known, 
the transition dipole distributions or orientations of the involved inter- and intraband transitions can be calculated from the dipole matrix elements.\cite{Heckmann2017}
 As shown there, intraband transitions within a given quantum well subband (e.g. $n_z=1$) have 100\% in-plane transition dipole orientation. 
This also holds for HH to CB transitions, while LH and SOH to CB transitions have both in- and out-of-plane dipole components. 
Calculating the resulting expected dipole distribution, we obtain 100\% in-plane orientation for paths (1) and (2), what excludes them also from likely paths. 
For paths (3) and (4), instead, 100\% in-plane and 66\% in-plane orientation are predicted, respectively. 
If both paths contribute equally, a 83\% in plane orientation of the absorption dipoles is expected, which is in good correspondence to the measured 85\%. 
This further supports the interpretation, that (3) and (4) are the most likely paths and responsible for the observed near quadratic area scaling of the TPA cross sections of CdSe nanoplatelets.

We infer from Table \ref{tab2} that the quadratic size scaling of $W_{TPA}$ is not exclusive of CdSe NPLs, but can be also found in other nanostructures with at least one ($N \geq 1$) weakly confined directions. These implies exceptionally high TPA cross sections can be obtained in quasi-1D, quasi-2D and quasi-3D systems, which are of immediate interest to all two-photon absorber applications, for instance for two-photon pumped lasing\cite{Li2015}, in confocal two-photon microscopy or TPA autocorrelation. We will reason the last example in the following.

Using Boyd\cite{BoydNLO} and Rumi et al.\cite{Rumi2010} we compare the efficiency of a standard BBO crystal second harmonic generation (SHG) autocorrelation and  that of a dense CdSe platelet TPA medium. For example we consider the autocorrelation of Ti:Sa Lasers with a  100\,MHz repetition rate and 100 fs pulses and a typical peak irradiance of 10 GW/cm$^2$ of a strongly  attenuated focussed beam (corresponding to a few mW power). For an interaction length of 10 micron, TPA has a 13\% efficiency compared to 5$\cdot 10^{-2}$\% for a BBO at 800\,nm. Even for an extremely short interaction length of 100\,nm the TPA autocorrelation exhibits a considerable 0.2\% efficiency compared to the vanishing 5$\cdot 10^{-6}$\% conversion efficiency of a BBO. Hence, TPA autocorrelation with nanoplatelets is far more efficient and allows such ultra short interaction lengths, that phase matching is not relevant any more. This further implies, that there are no relevant restrictions to the spectral bandwidth and pulse width of the signal to be correlated (like in BBOs for instance), since group velocity dispersion and mismatch do not limit the temporal resolution for these short interaction lengths. Hence it allows to measure high bandwidth and ultrashort pulses with superior sensitivity.

\section{Conclusions}

Our study has shown that radiative transition rates in semiconductor nanostructures with at least one weakly confined direction are very sensitive to excitonic correlation. If the transition is between strongly bound and weakly or unbound states, a pronounced dependence on the size of the particle is introduced. In particular, intraband and inter-VB transition rates decrease with increasing size.
The valence-to-conduction band transition rate can be enhanced by a factor $(L/a_B)^{N}$, with $L$ the length of the weakly confined direction, while that of intraband and inter-valence-band transitions can be slowed down by the inverse factor, $(a_B/L)^{N}$. This is the inverse of the GOST effect reported for interband transitions.

The gained principle understanding of radiative transitions in nanoscopic systems adds a new degree of freedom for the rational design of optical systems with highly tunable transition rates. Potential applications of our approach are two-photon autocorrelation and cross correlation with much higher sensitivity and unprecedented temporal resolution as well as TPA based optical stabilization, low excitation intensity confocal two-photon imaging and optimization of inter subband transition rates in quantum cascade lasers.

As a prominent example for the concept, we have shown the relevance for TPA processes, where exciton correlation of intermediate and final states provides a superlinear scaling of the transition rate with the nanostructure size. This offers an interpretation of the near quadratic volume scaling of two-photon absorption cross-section in CdSe nanoplatelets in line with results on the transition dipole orientation from two photon $k$-space spectroscopy. Thus  it lays foundations for the design of two-photon absorbers with outstanding performance and TPA cross sections well above those of conventional two-photon absorbers. Our approach can be extended to other nano materials including other II-VI nanocrystals, perovskites and transition metal dichalcogenides of different dimensionality. Further, our concept explains the background of the validity of the universal linear absorption continuum approach for the determination of particle concentrations via the intrinsic absorption.

\section{Methods}

\subsection{Effective mass model}

Electron and hole wave functions are described within a single-band k$\cdot$p model: $|j \rangle = | f_j \rangle\, | u_j \rangle$, with $f_j$ the envelope function and $u_j$ the periodic Bloch function.\cite{Voon2009} A dipole transition matrix element is then given by:
\begin{equation}
\langle i | H' | f \rangle = -\frac{e}{m}\,\mathbf{A}\,
\left( \langle u_i | \mathbf{p} | u_f \rangle \langle f_i | f_f \rangle + 
       \langle f_i | \mathbf{p} | f_f \rangle \langle u_i | u_f \rangle \right). 
\label{eq:dipole}
\end{equation}

\noindent where we have considered that the radiation-matter interaction Hamiltonian is $H'=-\frac{e}{m}\,\mathbf{A} \cdot \mathbf{p}$, with $e$ and $m$ the electron charge and mass. For nearly monochromatic light $\mathbf{A} = \frac{i}{\omega}\,\mathbf{E}$ is valid, with the local electric field $\mathbf{E}$. In Eq.\,(\ref{eq:dipole}) the first summand accounts for interband transitions ($u_i \neq u_f$) and the second one for intraband transitions ($u_i = u_f$), displayed schematically in Figure\,\ref{fig1}\,(b). Because $\langle u_i | \mathbf{p} | u_f \rangle$ relates to the Kane parameter (which depends on the material but not on the confinement) and $\langle u_i | u_f \rangle = \delta_{i,f}$, interband transitions are simply proportional to the overlap of the initial and final state envelope functions, $\langle f_i | f_f \rangle$, and intraband ones to  $\langle f_i |\mathbf{p}| f_f \rangle$. In the former case, the selection rule implies that $|f_i\rangle$ and $|f_f\rangle$ have the same envelope point symmetry.
In the latter case, the selection rule is less strict because $|f_j\rangle$ are not eigenfunctions of $\mathbf{p}$. Yet, the odd parity of the momentum operator implies that intraband transitions can only take place between initial and final states with different envelope function parity. We consider cuboidal nanostructures with infinite confinement potential and the related envelope functions for correlated and uncorrelated (IP) states (Figure\,\ref{fig1}\,(c)).

\subsection{Non-linear and linear absorption of CdSe NPLs}

By the means of open aperture z-scan\cite{Sheik-Bahae1990} and two-photon photoluminescence excitation (2P-PLE) spectroscopy at $800$\,nm we investigated the TPA cross sections of CdSe NPLs with varying lateral size.\cite{Scott2015} The setup and experimental conditions are described in Ref.\,\citenum{Scott2015}. CdSe NPLs with 3.5, 4.5 and 5.5 monolayer (ML) thicknesses were synthesized as described in Refs.\,\citenum{Achtstein14, Grim2014}, precipitated with methanol, redispersed in chloroform and inserted in 1\,mm fused silica cuvettes for spectroscopy. Lateral sizes varying from $5 \times 5\,$nm$^2$ to $82 \times 22\,$nm$^2$ were characterized by transmission electron microscopy.\cite{Scott2015}


\begin{acknowledgement}
J.P. and J.I.C. acknowledge support from MINECO project CTQ2017-83781-P and UJI project B2017-59.
R.S., N.O. and A.W.A acknowledge DFG grants WO477-1/32 and AC290-1/1 and 2/1. 
\end{acknowledgement}

\bibliography{Ref1}

\end{document}